\documentclass[namedreferences]{SolarPhysics}
\usepackage[optionalrh]{spr-sola-addons} % For Solar Physics 
\usepackage{epsfig}                     % For eps figures, old commands
\usepackage{graphicx}                    % For eps figures, newer & more powerfull
\usepackage{color}                       % For color text: \color command
\usepackage{url}                         % For breaking URLs easily trough lines
\begin{document}

\begin{article}

\begin{opening}

\title{Long-term Spatial and Temporal Variations of Aurora Borealis Events in the Period 1700--1905}

%%%%%%%%%%%%%%%%%%%%%%%%%%%%%%%%%%%%%%%%%%%%%%%%%%%
%% Authors Names
%
\author{M.~\surname{V\'azquez}$^{1,2}$\sep  J.M.~\surname{Vaquero}$^{3}$\sep  M.C.~\surname{Gallego}$^{4}$}

%%%%%%%%%%%%%%%%%%%%%%%%%%%%%%%%%%%%%%%%%%%%%%%%%%%
%% Runningheads
%
\runningauthor{V\'azquez, Vaquero, and Gallego}
\runningtitle{Aurora Borealis}

%%%%%%%%%%%%%%%%%%%%%%%%%%%%%%%%%%%%%%%%%%%%%%%%%%%
%% Affilations 
%
  \institute{$^{1}$ Instituto de Astrof\'\i sica de Canarias, 38200 La Laguna, Spain.
                     email: \url{mva@iac.es} \\
            $^{2}$ Departamento de Astrof\'\i sica, Universidad de La Laguna, 38205 La Laguna, Spain.\\
            $^{3}$ Departamento de F\'\i sica, Universidad de Extremadura, Avda. Santa Teresa de Jornet 38, 06800, M\'erida, Spain.
                     email: \url{jvaquero@unex.es}\\
             $^{4}$ Departamento de F\'\i sica, Universidad de Extremadura, 06071  Badajoz, Spain.
                     email: \url{maricruz@unex.es}\\
            }

%%%%%%%%%%%%%%%%%%%%%%%%%%%%%%%%%%%%%%%%%%%%%%%%%%%

\begin{abstract}
  Catalogues and other records of aurora-borealis events were used
  to study the long-term spatial and temporal variation of
  these phenomena in the period from 1700 to 1905 in the
  Northern Hemisphere. For this purpose, geographic and
  geomagnetic coordinates were assigned to
  approximately 27\thinspace000 auroral events with more than 80\thinspace000
  observations. They were analysed separately in three
  large-scale areas: i) Europe and North Africa, ii)
  North America, and iii) Asia. There was a clear need to
  fill some gaps existing in the records so as to have a reliable
  proxy of solar activity, especially during the 18$^{th}$
  century. In order to enhance the long-term variability, an
  11-year smoothing window was applied to the data. Variations in the cumulative numbers of auroral events
  with latitude (in both geographic and geomagnetic
  coordinates) were used to discriminate between the two
  main solar sources: coronal mass ejections and high-speed
  streams from coronal holes.  The
  characteristics of the associated aurorae
  correlate differently with the solar-activity cycle.
\end{abstract}

%%%%%%%%%%%%%%%%%%%%%%%%%%%%%%%%%%%%%%%%%%%%%%%%%%%
%

\end{opening}
%-------------------------------------------------

\section {Introduction}

Aurorae are the most easily seen manifestations of the
interaction (compression and magnetic reconnection) between
the solar wind and the Earth's magnetosphere. Just
naked-eye observations are sufficient, without any need for specific
training. They are therefore a useful tool with which to study the
variability of this process in  past centuries because
several catalogues, books, and reports of various kinds are
available. Here we shall concentrate on two recent centuries. For
longer periods of time (thousands of years), the use of
cosmogenic isotopes, such as $^{14}$C and $^{10}$Be, or of the nitrate
content in polar ice is recommended (Usoskin and
Kovaltsov, 2012; Traversi {\it et al.}, 2012).

The visibility of aurorae is limited to ring-shaped regions
around the geomagnetic poles -- the auroral ovals, centred
at around 65 degrees magnetic latitude in each hemisphere.
Under the impact of a solar storm, the auroral ovals undergo
broadening, particularly on the night side.  Low-latitude
aurorae are very rare, and are clearly associated with strong
geomagnetic storms produced by solar coronal mass ejections.
They are generally red and diffuse, resulting primarily from
an enhancement of the 630.0 nm [OI] emission due to
bombardment by soft electrons ($<$100 eV) precipitating from
the plasmasphere (Tinsley {\it et al.}, 1986). The typical altitude for a
low-latitude aurora is 250\,--\,400 km (Roach {\it et al.},
1960). They can be confused with distant fires or twilight
due to their colour.  The limits of such aurorae are usually
15 and 45 degrees geomagnetic latitude.  A similar
phenomenon, which shares the same energy source, is that of
stable auroral red arcs. They are observed mainly during the
recovery phase of geomagnetic storms (Rassoul {\it et al.},
1993; Nakazawa, Okada and Shiokawa, 2004).

The  visual sensitivity thresholds  of the green and red
radiation of the aurorae are between one and ten kilorayleighs,
which means that only those aurorae coming from moderate and
strong geomagnetic storms will be visible and therefore part
of our sample (Schr\"oder, Shefov, and Treder, 2004). It is clear that the
meteorological conditions also play a role, although a
cloud-free sky is not necessary. According to Livesey (1991),
the zone with the greatest aurora-borealis probability passes
across northern Norway, over to Iceland, south of Greenland,
and over to the South of Hudson Bay in North America.

Fritz (1873)  produced the first graph showing the geographical
distribution of auroral frequencies, measured in nights per
year. The values ranged from a minimum in the Mediterranean area
to a maximum at a magnetic latitude of 67 degrees.

In previous articles (V\'azquez {\it et al.}, 2006; V\'azquez
and Vaquero, 2010), we have studied low-latitude auroral
events. We established a correlation with several indices of
solar activity and drifts of magnetic latitude at two
different sites,  verifying the adequacy of such
terrestrial events to track the long-term variation of solar
activity in the past.  For a monograph on historical aspects
of solar activity, see Vaquero and V\'azquez (2009).

In the present work, we considered the period 1700\,--\,1905 for
which various documental sources of auroral observations are
available, indicating the date and place of the
observations.  This allowed geographic coordinates to be
ascribed to each observation, with a maximum of six sites
being taken per day and per catalogue.  If necessary, a
selection was made from the existing sample to give the
broadest coverage in both longitude and latitude. For a
previous work on auroral activity in this period, see Feyman
and Silverman (1980), although their study is restricted
spatially to Scandinavian and New England sources.

\begin{figure}[ht]
 \centerline{\includegraphics[width=0.95\textwidth,clip=]{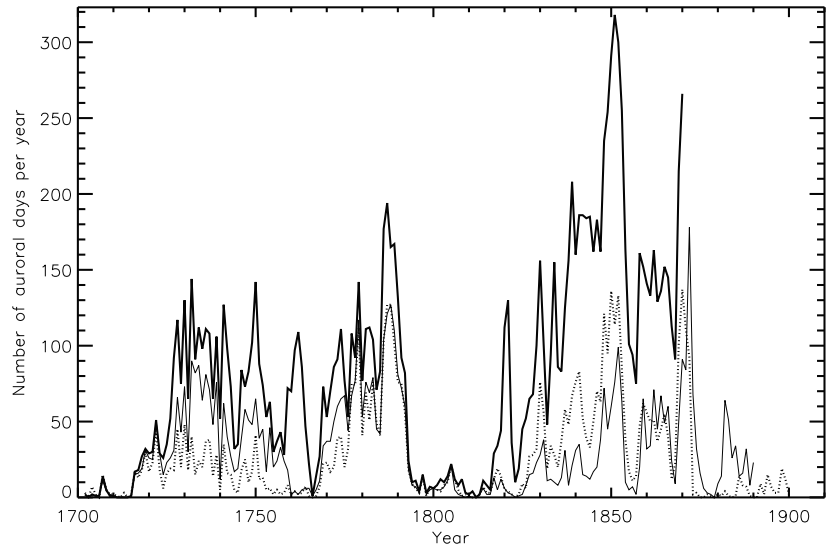}}
\caption{Yearly number of aurorae in the following catalogues: (dotted line) Krivsk\'y and Pejml (1988) updated by 
 Krivsk\'y (1996), (thin solid line) Angot (1897), and (thick solid line) Fritz (1873). 
 The first two catalogues are limited to zones further south than 55 degrees of geographic latitude.}
\label{aurkr}
\end{figure}

Table \ref {aurCat} lists the documentary sources used in
the present work. The number of days in the period studied
(1700\,--\,1905) was approximately 73\thinspace000 (200 years).  The available
data for the Northern Hemisphere were divided into three
continental areas corresponding to Europe (and North
Africa), North America, and Asia. The main sources of data
were the catalogue of Fritz (1873) and the archives of S.
Silverman\footnote{These include: (a) Catalog of Ancient Auroral Observations, 666 BCE to 1951; 
(b) The Auroral Notations from the Canadian Monthly Weather Review; 
(c) The New England Auroral Observations (1720\,--\,1998);  and (d) Daily Auroral Reports, 
Southeastern Canada and Northeastern US (1848\,--\,1853).}. For only a few places (less than ten) was it  impossible
to unequivocally determine the site's coordinates.  More
frequently, only the names of countries or US states were
described, and in these cases average coordinates were
assigned.

Fritz's North American data are mainly based on the earlier
catalogues of Lovering (1866) and Loomis (1860, 1861). They
are complemented for the later phase of the 19$^{th}$ century with
data from Greeley (1881) and different archives collected by
S. Silverman.

For Europe, the Fritz data were complemented with additional
values for high latitudes from Rubenson (1882) and
Tromholt\footnote{For a biography of S. Tromholt see Moss and
 Stauning (2012).} (1898). Particularly relevant are the Greenland
observations, contained in the Fritz catalogue and the
Silverman archives (see Stauning, 2011). For the rare low
latitude observations, we included data for the Iberian
Peninsula\footnote{They include published data from Vaquero, Gallego, and Garcia
(2003), Vaquero and Trigo (2005), Aragon\`es Valls
  and Orgaz Gargallo (2010), and Vaquero {\it et al.} (2010).} and
the Canary Islands (V\'azquez and Vaquero 2010).

 \begin{table}
 \caption{List of catalogues used for this study. The S. Silverman archives are available from the
National Space Data Center (http://nssdcftp.gsfc.nasa.gov/miscellaneous/aurora/).}
 \label{aurCat}
\begin{tabular}{llrr}
\hline
Observing Area & Catalogue &  Number of  & Number of \\
  &  & auroral days & locations\\
\hline
Europe and N. Africa & TOTAL & 25\thinspace306& 43\thinspace460 \\
       & Fritz (1873) & 9761 & 15\thinspace874 \\
       & S. Silverman Archives & 6089 & 11\thinspace810\\
       & Angot (1897)   & 439 & 753\\
	& Rubenson (1892) & 5924 & 9809 \\
        & Tromholt (1902) & 2311  & 4328\\
        & Sunderland Archive & 453 & 454\\
        & Hungary (R\'ethly and Berkes, 1963)   & 97 & 138 \\
        & Visser (1942) & 93 & 93\\
        & Harrison (2005) & 8 & 8 \\
        & W. Schr\"oder (1966) & 3  & 5\\
        &  Iberian Peninsula & 76 & 109\\
        & V\'azquez and Vaquero (2010) & 9 & 12 \\ 
        & Different Journals  & 8  & 12  \\
\hline
North America  &  TOTAL & 15\thinspace617 & 38\thinspace275\\
             & Fritz (1873) & 5351 & 11\thinspace160\\
            & S. Silverman Archives & 8605 & 23\thinspace674\\
	    & Greeley (1881) & 1379 & 2919\\
	    & Lueders (1984) & 121 & 121 \\
            & Broughton (2002) & 14 & 14\\
             & Mendillo and Keady (1976) & 5 & 5\\
	   & Eather (1980) & 4 & 4 \\	  
\hline
Asia   & TOTAL & 324 & 359\\
     & Fritz (1873) & 245 & 270 \\
     & Yau et al. (1995) & 27 & 27 \\
      & Lee et al. (2004) & 6& 8\\
     & Willis, Stephenson and Fang (2007) & 33 & 36\\
      & S. Silverman Archives & 11 & 14\\
      & Basurah (2004) & 2 & 2 \\
          & Harrison (2005) & 1 & 1 \\
\hline
\end{tabular}
\end{table}

The data sample is clearly inhomogeneous in both space and
time. This reflects not only meteorological variations but also the
difficulties of access of some regions. For North America,
there are many temporal gaps until 1746, and, after that, no
data are available for the years 1754, 1755, 1756, 1766,
1799, 1810, or 1812. There is a remarkable contribution for
the entire period studied from numerous forts (more than 60)
located along the borders of the expanding settlement of the
western and northern territories.

After the Maunder Minimum, the works of Halley (1716) and
Mairan (1733) marked the beginning of modern studies of
aurorae.  Krivsk\'y and Pejml (1988) and Krivsk\'y (1996)
compiled a catalogue (archived at the National
  Geophysical Data Centre) recording the aurorae visible in
Europe in the period 1715\,--\,1850 at latitudes lower than 55
degrees. They assumed that nearly all of the aurorae were
recorded after roughly 1720, and that therefore no
normalization factor needed to be applied (Figure \ref
{aurkr}). These records are compared with the catalogue of
Angot (1897) for the period 1700\,--\,1890, also covering
observations at latitudes lower than 55 degrees, and Fritz
(1873) with no restrictions on latitude. Unfortunately, no
such global catalogues exist for the 20$^{th}$ century.

The reduction in the number of aurorae observed during the
Dalton Minimum is confirmed (Silverman, 1992; Broughton,
2002; Vaquero {\it et al.}, 2003). Two strong episodes of
auroral activity are clearly seen around 1775 and 1850
(Figure 1). Around the turn of the 19$^{th}$ to the 20$^{th}$ century, the level of
solar activity decreases, as also does the visual-aurora
monitoring. There is no recovery from this decrease until
the start of the space age in 1957 (Siscoe, 1980; Legrand
and Simon, 1987; Silverman, 1992).

It is our purpose now to proceed to a careful statistical
study of these classical catalogues, expanding them with
other available sources. Our main emphasis is on providing a
3D map (geographic coordinates and time) of the auroral
activity in the past, when auroral observations constitute
the only information available about heliosphere activity.

\begin{figure}[ht]
 \centerline{\includegraphics[width=0.95\textwidth,clip=]{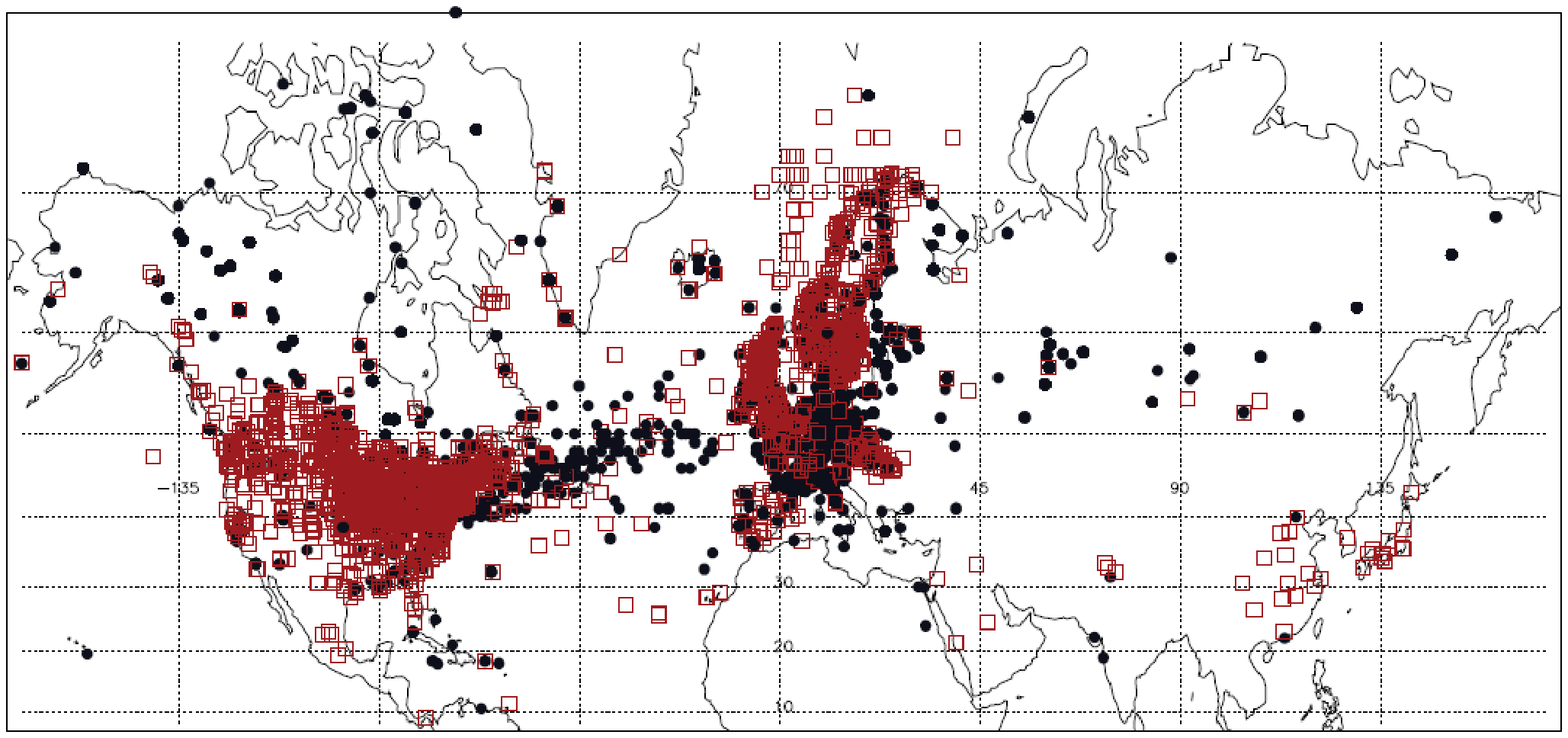}}
\caption{Places where aurorae were visible at least once in the Northern Hemisphere during the 
studied period 1700\,--\,1905: ($\bullet$) Fritz and Angot catalogues; (red squares) other catalogues and reports.}
\label{GlMap}
\end{figure}

Figure \ref {GlMap} shows the locations of auroral
observations on a Mercator projection of the Northern
Hemisphere. The accompanying Tables \ref {aurEurCity} and
\ref {aurAmCity} indicate the sites where most auroral
events were reported for Europe and North America
respectively. The numbers are clearly low for Eastern Europe
and Asia.

\begin{table}
 \caption{Sites where aurorae were most frequently observed in Europe.
The geographical coordinates are expressed in degrees. ($\ast$) Referred in the catalogues 
only to the country, without assigning any defined site.}
 \label{aurEurCity}
\begin{tabular}{llrr}
\hline
Site & G. Latitude & G. Longitude & Auroral events\\
\hline
Godhaab (Nuuk) & 64.175 & -51.739 & 2164\\
Uppsala  & 59.85 & 17.63 & 1973 \\
Stockholm & 59.35 & 18.08 & 1593\\
Christiania (Oslo)  & 59.92 & 10.76 & 1536 \\
Jakobshavn (Ilullissat) & 69.17 & -49.83 & 1487\\
Ofver-Tornea & 66.38 & 23.65 & 1016\\
Saint Petersburg & 59.95 & 30.32 & 968\\
Scotland ($\ast$)  & 56.00 & -4.00 & 951\\
Utrecht & 52.09 & 5.12 & 818\\
H\"arnosand  &  62.66   &  17.94& 813\\
Berlin & 52.52 & 13.41 & 623 \\
Paris & 48.86 & 2.35 & 464\\
Lund  &  55.70   &  13.20& 427\\
\hline
\end {tabular}
\end {table}

\begin{table}
 \caption{Sites where aurorae were most frequently observed in North America.
The geographical coordinates are expressed in degrees.}
 \label{aurAmCity}
\begin{tabular}{llrr}
\hline
Site & G. Latitude & G. Longitude & Auroral events\\
\hline
Toronto  & 43.72 & -79.34 & 1800 \\
Cambridge, MA &42.37 & -71.11 & 921\\
New Haven  & 41.31 & -72.92 & 908 \\
Gardiner, ME &  44.21 & -69.79 & 798 \\
New York City and State & 43.00 & -75.00 & 734\\
Winnipeg & 49.90 & -97.14 & 727\\
York Factory &  57.01   &  -92.31   & 518 \\
Quebec  & 46.82 & -71.22 & 494 \\
Eastport  & 44.91 & -67.00 & 486 \\
Burlington & 44.47 & -73.15 & 340 \\
Moose Factory & 51.26 & -80.59 & 287\\
Montreal  &  45.51 & -73.55 & 259\\
Point Barrow & 71.39 & -156.48 & 251 \\
\hline
\end {tabular}
\end {table}

\begin{figure}[ht]
 \centerline{\includegraphics[width=0.95\textwidth,clip=]{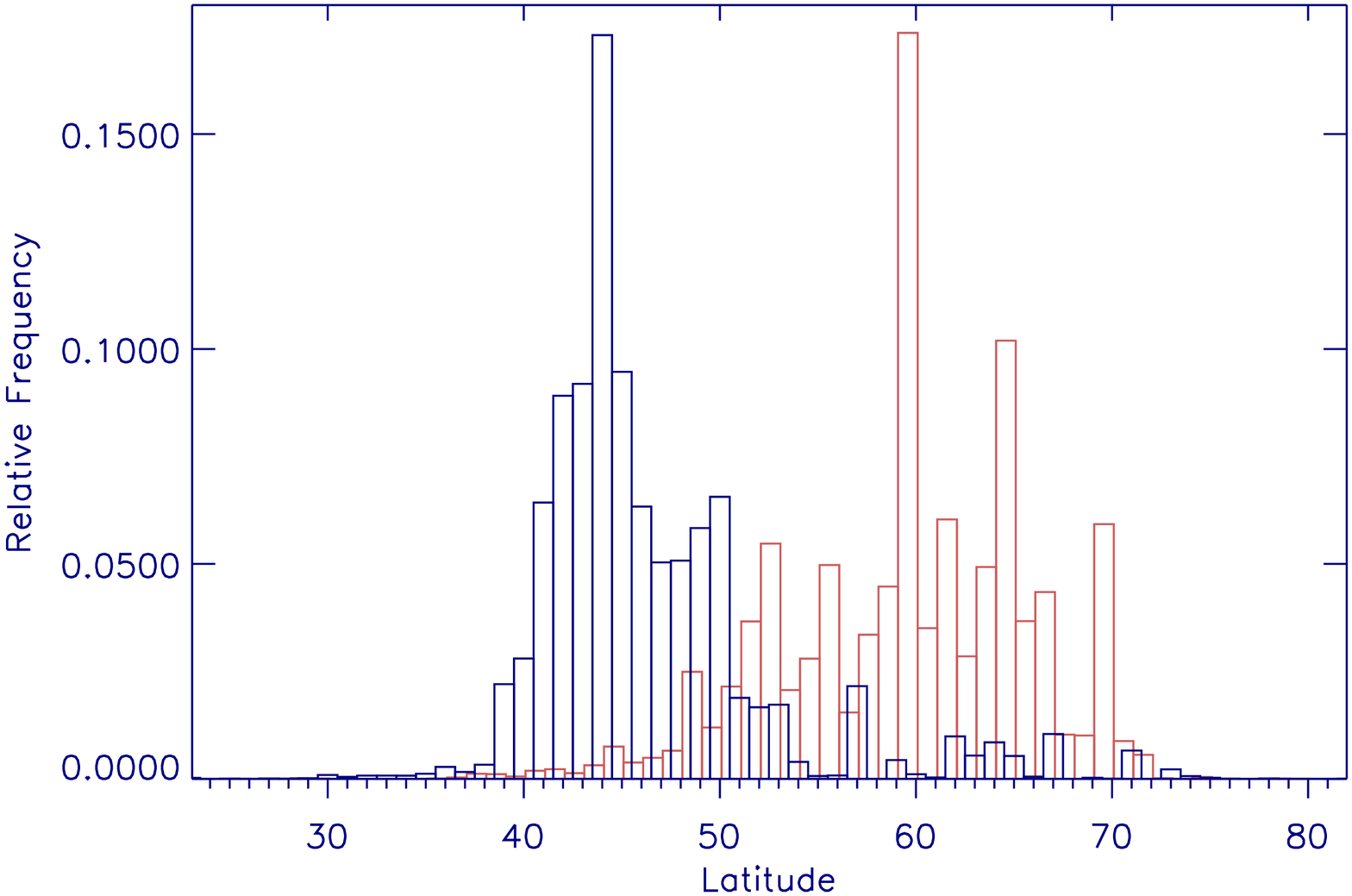}}
\caption{Distribution of geographic latitudes for Europe (red) and North America (blue) using the different catalogues.}
\label{Hist}
\end{figure}

Figure \ref {Hist} shows the histograms of the latitude
distributions for Europe and North America. They both
suggest a bimodal distribution, with the maxima at middle
and high latitudes for North America and Europe,
respectively. The expected trend would be an increase of
auroral events at high latitudes, but in the case of North
America the histograms reflect the slow settlement of the
northern areas.  Moreover, access to southern data was
delayed until these regions (Louisiana, Florida, Texas, New
Mexico) were annexed by the USA.  A search of 18$^{th}$ century sources in Spanish would be worthwhile
for this region.

We checked the relationship between the auroral observations
in these catalogues and moonlight, since an aurora should be
easier to observe when there is little moonlight during the
night. In particular, the light of the full Moon would
impede the observation of weak aurorae. Therefore, we
expected relatively more (fewer) reported aurorae when there
was a new (full) Moon. Indeed, we found that there was a
clear decrease of observed events with increasing brightness of the moon, as was
expected.

\section {Temporal Variation}

The number of aurorae is clearly related to the level of solar activity. 
The records reflect the well-known Sp\"orer, Maunder,
Dalton, and 1901\,--\,1913 minima, and include a previously unrecognized long-term
minimum around 1765 (Silverman, 1992, 1995). 

A double peak in the geomagnetic records has been observed,
with the  two maxima occurring i) shortly before the sunspot maximum
and produced by transient events (Gonzalez {\it et al.},
2002), and ii) two years after the maximum and mainly produced by
recurrent events (Tsuratani {\it et al.}, 2006). Geomagnetic
activity is greater in the second half of even-numbered
cycles and in the first half of the odd-numbered cycles,
giving rise to a 22-year variability (Chernosky, 1966:
Vennerstr$\o$n and Friis-Christensen, 1996).  For a
discussion of the geoeffectiveness of the different solar
sources of geomagnetic disturbances, see Georgieva {\it et
al.} (2006).

\begin{figure}[ht]
 \centerline{\includegraphics[width=0.85\textwidth,clip=]{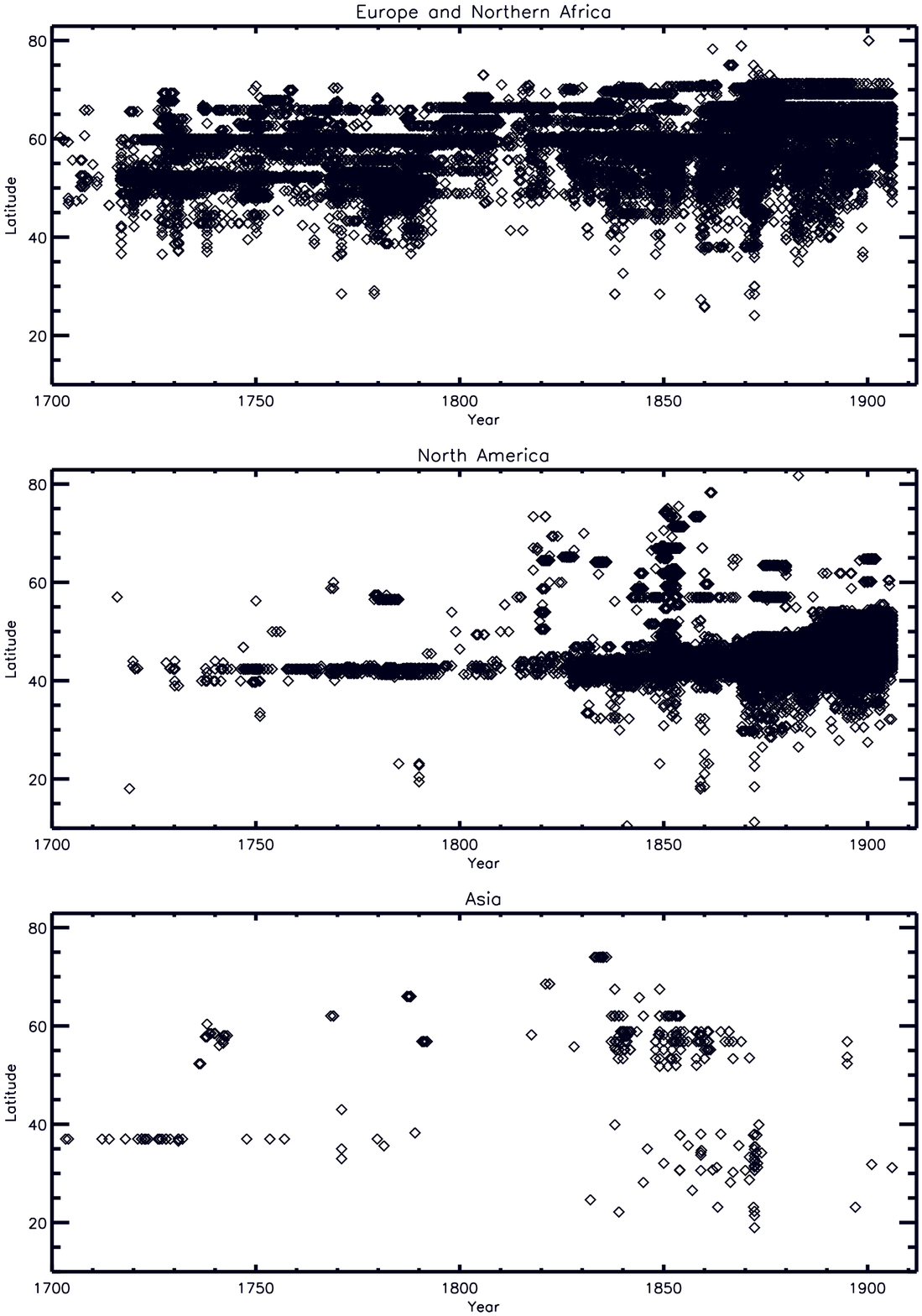}}
\caption{Temporal and latitudinal variation of aurora borealis events visible during the 
18th and 19th centuries from different archives (see Table 1) in (top) Europe and North Africa, 
(middle) North America, and (bottom) Asia.}
\label{GloTime}
\end{figure}

The variation in time and latitude of the auroral
observations recorded in our sample is shown in Figure \ref
{GloTime}.  The European data reflect the Dalton Minimum
very clearly. The lack of high-latitude records in the Fritz
catalogue was remedied by including the Scandinavian
observations of Tromholt and Rubenson. The aurorae observed
in Barcelona in 1811 and 1812 are remarkable, and call for
detailed confirmation. The existence of the secondary
Silverman Minimum (around 1760) seems also to be confirmed.
The middle panel of the figure reflects the paucity of North
American data during the 18$^{th}$ century, especially at
high latitudes.  The settlement of the western regions
produces a clear increase in the number of
reports.  The records in New England are,
however, relatively frequent over the whole period.
Finally, the bottom panel reflects the scarcity of the data
that we were able to collect for auroral observations in
Asia.  In light of this, in the analyses below we shall
concentrate  mainly on European and American
sources. The combination of the different auroral catalogues
shows a quasi-80-year periodicity (the Gleissberg cycle), confirming
previous findings ({\it e.g}. Siscoe, 1980).

The primary aim of this article was to study the long-term auroral variability gathering together different 
sources of auroral information. In the following we will calculate some correlations between sunspot numbers 
and different auroral parameters. In order to enhance this long-term variability an 11-year smoothing window 
will be applied to the data sets. Clearly, this procedure will increase the strength of the correlation, 
because we are entering a frequency common to the two sets, and will decrease the 
degrees of freedom of the series (number of independent points). To evaluate the statistical significance of 
the correlation we have applied the {\it t}-student test, calculating the significance given by the {\it p}-value. 
To take into account the effect of the smoothing in the calculation of the {\it t}- and {\it p}-parameters,
we have reduced the number of independendent points by a factor of 11, the length of our smoothing window. 
We have verified that the effect of the smoothing on the correlations is very small when the length of the 
smoothing window is smaller than the length of the data set. For our sample this criteria
is fullfilled when the number of available data is larger than $\approx$ 140 years. Finally, we can point out  
that the correction to the number of statistically independent points resulting from autocorrelation is negligible.

\section {Latitudinal Variation}

\subsection {Solar Source of the Aurorae}

One can divide the effects of solar activity on our planet
via the solar wind as being due to  two agents, depending on
the topology of the magnetic field of the solar atmosphere (see
Legrand and Simon, 1981).

i) {\it Closed Regions}: The magnetic flux of the active
regions is characterized by magnetic configurations with
closed field lines dominating the variations in the total
irradiance and emission in the high energy range of the
solar spectrum (ultraviolet and X-rays).  For historical
reasons, the sunspot number is usually employed as a measure
of this type of activity.  Coronal mass ejections (CMEs) are
transient phenomena linked to large-scale reorganizations of
the magnetic field.  They consist of huge emissions of solar
particles that in some cases impact the Earth's environment.
The CME rate closely follows the solar cycle (Webb and
Howard, 1994; Robbrecht, Berghmans, and Van der Linden, 2009) and the
Gnevyshev gap (Gnevyshev, 1967), but the main physical
parameters of CMEs lag the sunspot number by 19 months, at least for solar cycle 23 (Du, 2012).

ii) {\it Open regions}: Large-scale magnetic regions have
field lines open towards the interplanetary medium. They are
the main source of a continuous outward flow of charged
particles (protons, electrons, and helium nuclei) known as
the solar wind. Long-lived coronal holes are the sources of
high-speed solar wind, and are related to recurrent
geomagnetic activity. They occur more frequently in the
declining phase of the sunspot cycle (Verbanac {\it et al.},
2011).

The solar magnetic field of the open regions, the open
magnetic field (OMF), is frozen into this wind, thereby
configuring the interplanetary magnetic field (IMF) which
produces the heliosphere that fills practically the whole
Solar System. Lockwood {\it et al.} (1999) and Lockwood
(2003) derived the intensity of the interplanetary
magnetic field from the aa geomagnetic intensity index, a record
that extends back to 1868. They found that the average
strength of the solar magnetic field has doubled in the last
100 years.  Usoskin {\it et al.} (2002) carried out
simulations of the variations of the OMF with a model based
on the emergence and decay rates of active regions (Solanki, Sch\"ussler, and Fligge, 2002). 
That model fits reasonably well other
proxies of solar activity, such as the $^{10}$Be records in
ice cores.  Since solar cycle 14 (1902\,--\,1913), the geomagnetic
activity lagged behind the sunspot number, but before that date
the lag seems to have been less notable (Love, 2011).

Siscoe (1980) discriminated between aurorae that are visible
North and South of 54 degrees using Scandinavian records,
noting that the southern data tracked the 11-year solar cycle
more clearly.  Bravo and Otaola (1990) studied the location
of solar coronal holes and their influence on the auroral
records.  They found that the number of aurorae is positively
correlated with polar coronal holes that reach solar
latitudes below 60 degrees. Verbanac {\it et al.} (2011)
found that high-speed streams originating in equatorial
coronal holes are the main driver of solar activity in the
declining phase.

In order to differentiate between the distinct solar
sources, we plotted the cumulative values, ${\rm
N_{Lat,j}}$, of the number of aurorae visible for each
1.5 degree wide latitude band (see the definition below).

\begin {equation}
{\bf {\rm N_{Lat,j} = \Sigma_0^{i=j} N_{Lat,i}}}
\end {equation}

\begin{figure}[h]
 \centerline{\includegraphics[width=0.88\textwidth,clip=]{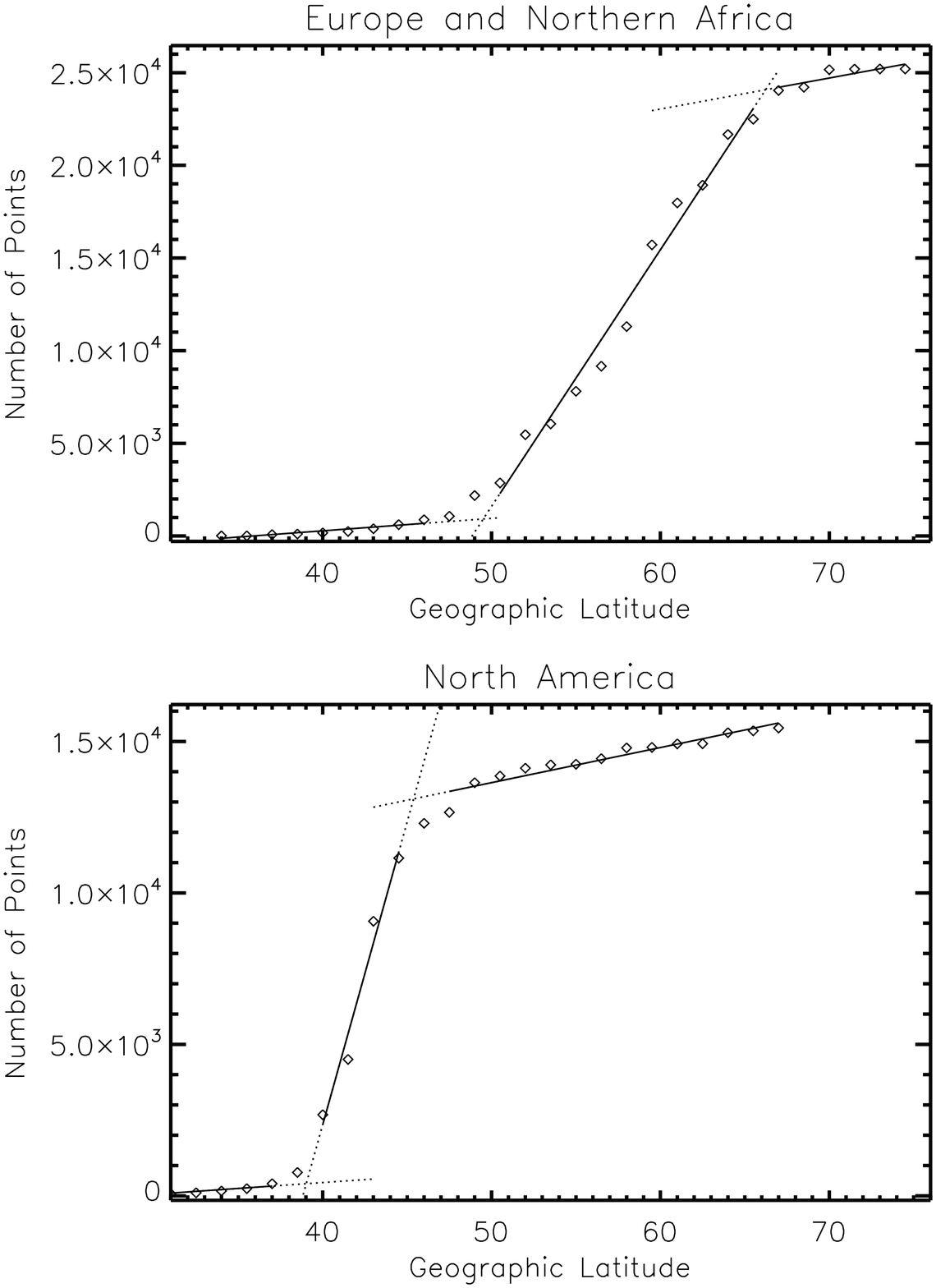}}
\caption{Cumulative values of the number of aurorae visible per 1.5 degrees latitude bin 
(see text for explanation). Top panel: Europe and North Africa; bottom panel: North America.}
\label{aurLatGlob}
\end{figure}

The growth found is almost exponential.  We took the
resulting curve to consist of three segments, to each of which
we made straight-line fits.  The intersection of the three
segments could represent the boundaries between different
solar sources of the auroral event. The two plots of Figure
\ref {aurLatGlob} show the results for Europe and North
America, respectively.

The low-latitude segment would represent the aurorae
produced by strong solar storms, with the northern limit at
49 degrees in Europe and 39 degrees in North America
defining the so-called low-latitude aurorae.  The middle
segment shows auroral events produced by CMEs of medium
strength.  In the upper latitude segment, the predominant
role is played by the fast streams from coronal holes.  In
this case, the variation of the geographic latitude plays
only a minor role. The limits are approximately 66 degrees
for Europe and 44 degrees for North America.

\subsection {Width of the Auroral Oval}

Viewed from space, aurora are diffuse ovals of light around
the geomagnetic poles. They can be regarded as the regions
of the upper atmosphere with permanent luminescence
(Feldstein, 1986).  All of the great auroral expansions have
been associated with intense values of the interplanetary
magnetic field (Sheeley and Howard, 1980). The variation in
the radius of the auroral oval in response to solar wind
changes has been studied by Milan {\it et al.} (2009).

Assuming a spherical Earth with a radius R = 6371 km, one
calculates the area (A) of a longitude--latitude rectangle
(area between two lines of latitude lat$_{\rm 1}$ and
lat$_{\rm 2}$) (Moritz, 1980), as

\begin {equation}
{\rm A = 2 \pi R^2 (\sin {lat_1} - \sin{lat_2}). }
\end {equation}

\begin{figure}[ht]
 \centerline{\includegraphics[width=0.96\textwidth,clip=]{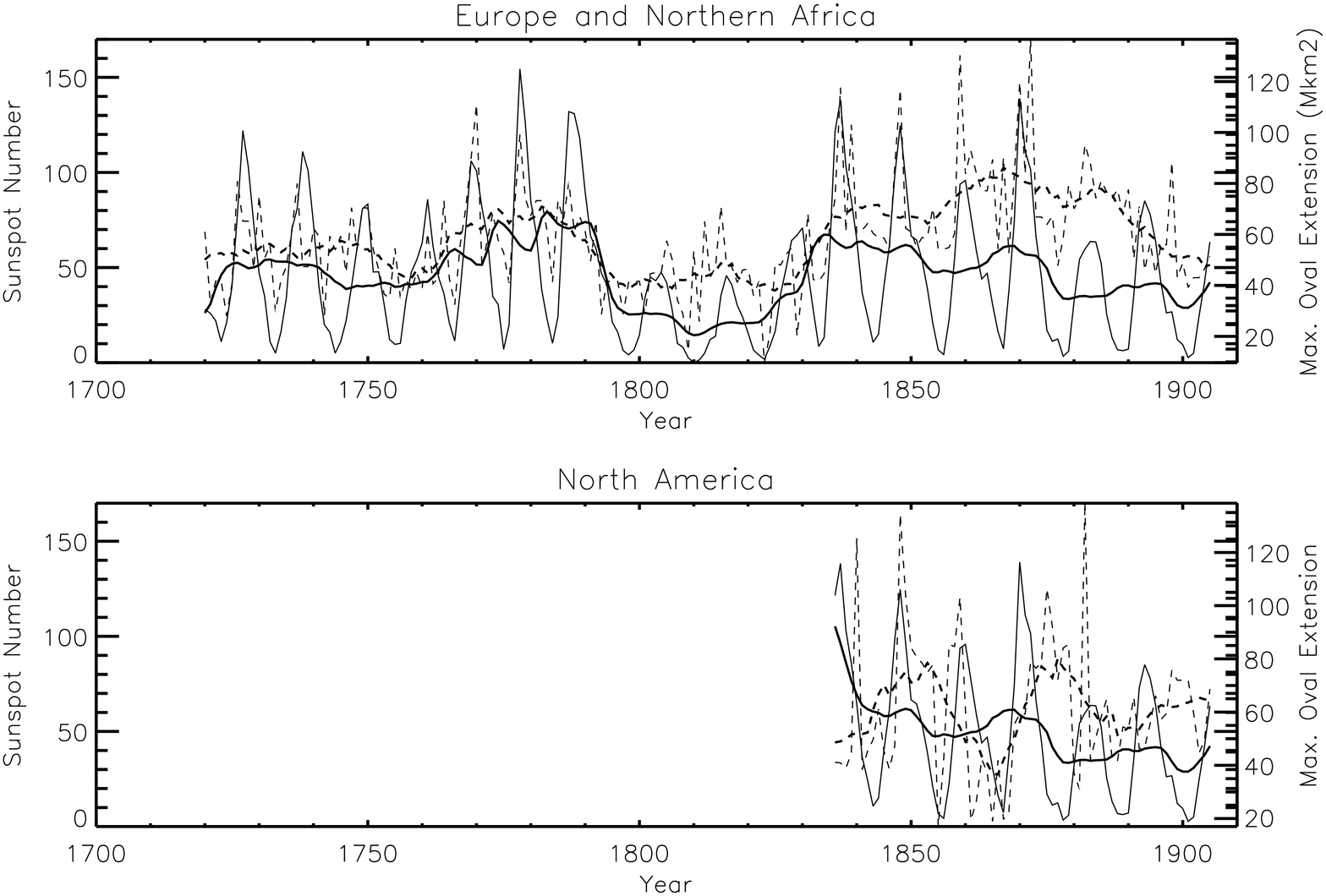}}
\caption{Variation of the annual maximum of the oval extension (dashed) with the annual sunspot number (solid lines). 
The thicker line corresponds to an 11-year smoothing. (Top panel) Europe and North Africa;
(bottom panel) North America. The area of the ovals is given in millions of square kilometres.}
\label{GloLatmaxy}
\end{figure}

Figure \ref {GloLatmaxy} shows the variation of the annual maximum of the oval extension with the 
annual sunspot number.  One must bear in mind
that the auroral oval is really centred on the geomagnetic
pole, so that the above expression is just a first-rough
approximation.  For the European data (sample 1720\,--\,1905)
the level of correlation is not very strong but significant (r=0.58, {\it p}-value $<$ 0.02). 
However, for the North American data (sample 1836\,--\,1905) the results are much
noisier, probably reflecting the shortness of the sample
with adequate data and the lack of observations of auroral
events at extreme latitudes for many years.

\subsection {Low-latitude Aurorae}

\begin{figure}[ht]
 \centerline{\includegraphics[width=0.90\textwidth,clip=]{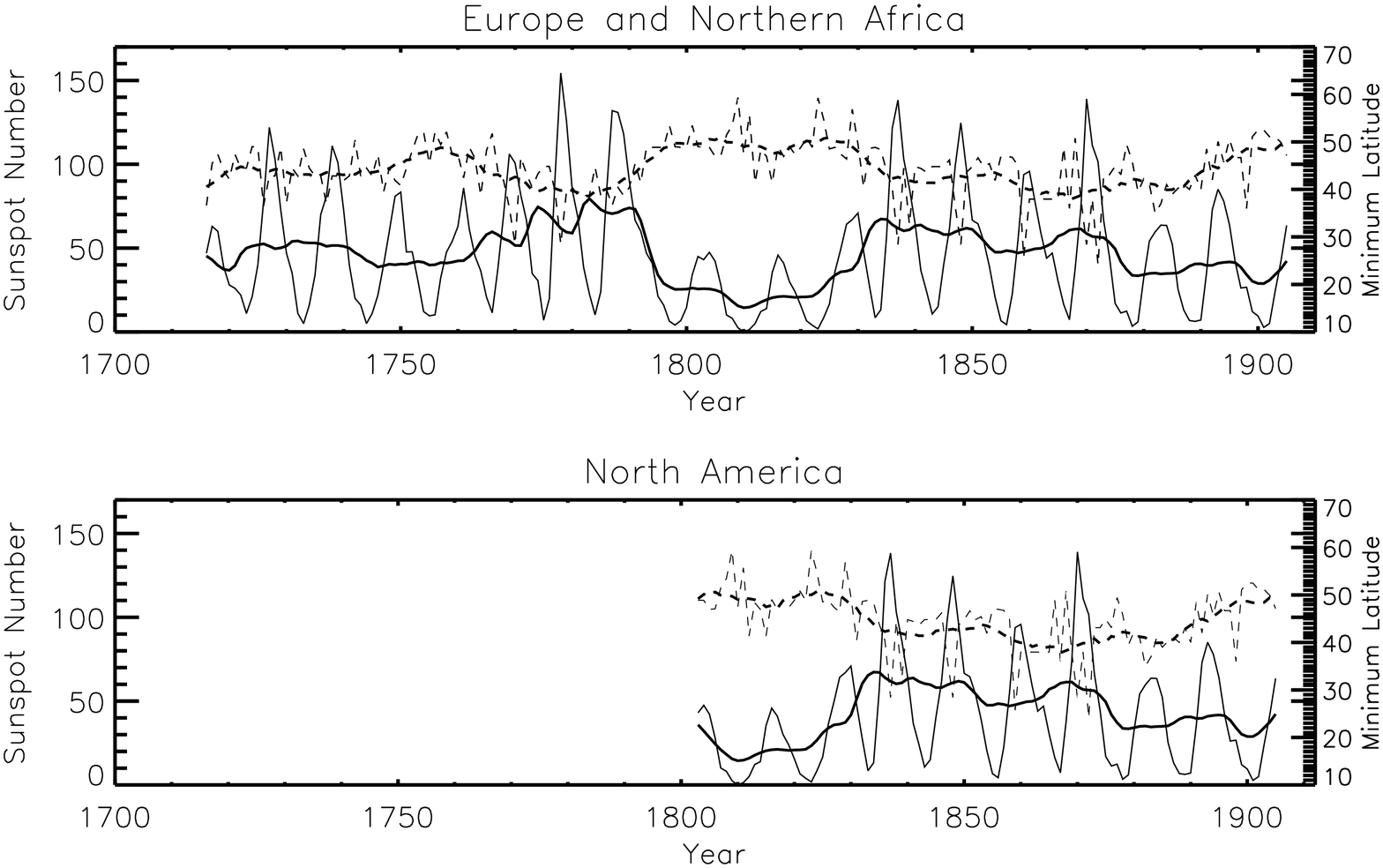}}
\caption{Variation of the annual minimum latitude (dashed, right axis) with the annual sunspot number (solid lines, 
left axis). The thicker line corresponds to an 11-year smoothing of the data. (Top panel) Europe and North Africa; 
(bottom panel) North America.}
\label{GloLat1y}
\end{figure}

A solar storm is accompanied by an expansion of the auroral oval towards the Equator on the night side of the Earth. 
Figure \ref {GloLat1y} shows the temporal variation of the yearly minimum latitude and its correlation 
with the sunspot number.  For the North American sample, we restricted our observing period to the data 
after 1803 due to the low latitudinal and temporal coverage at earlier dates. 
When the data were smoothed with an 11-year window, the correlations for both data samples were 
good ({\it r} = -0.74 and -0.71 for the European and the American data, reflecting the 80-year period of 
variability of solar activity. However, the level of significance was much better for the European 
({\it p}-value $<$ 0.001) than for the North American data ({\it p}-value $<$ 0.05), due to the longer data set.

The lowest latitudes were 24.09 degrees for Europe, 24.08 degrees for North America, and
18.97 degrees for Asia. All of these values correspond to the aurora
of  4 February 1872 during Solar Cycle 11 (see Silverman,
2008).

In this context, we should remark that low-latitude aurorae
have also been observed during periods of weak to moderate
geomagnetic activity.  Silverman (2003) has shown from US
auroral data from 1880 to 1940 that some of the auroral
phenomena occurred under conditions of quiet or moderate
magnetic activity and at low latitudes. He used the term
``sporadic aurora'' for this type of auroral phenomenon.
Vaquero, Trigo and Gallego (2007) studied one such event observed
in 1845 in the Iberian Peninsula. Willis, Stephenson, and Fang (2007)
compiled 42 Chinese and Japanese auroral observations during
the period 1840\,--\,1911 and found that at least 29 of the 42
observations ({\it i.e.} 69 \%) occurred at times of weak to
moderate geomagnetic activity. 

\subsection {High-latitude Aurorae}

The Greenland observations confirmed that, at high
latitudes, the aurorae maximum coincides with the sunspot
cycle minimum (Anonymous, 1883). In order to check
the reliability of our method in discriminating between
different solar sources, we looked for a critical latitude
at which this anti-correlation is largest.  For this
purpose, we normalized the data above a critical latitude to
the total number of aurora observations, and plotted the
resulting temporal variation.

\begin{figure}[ht]
 \centerline{\includegraphics[width=0.96\textwidth,clip=]{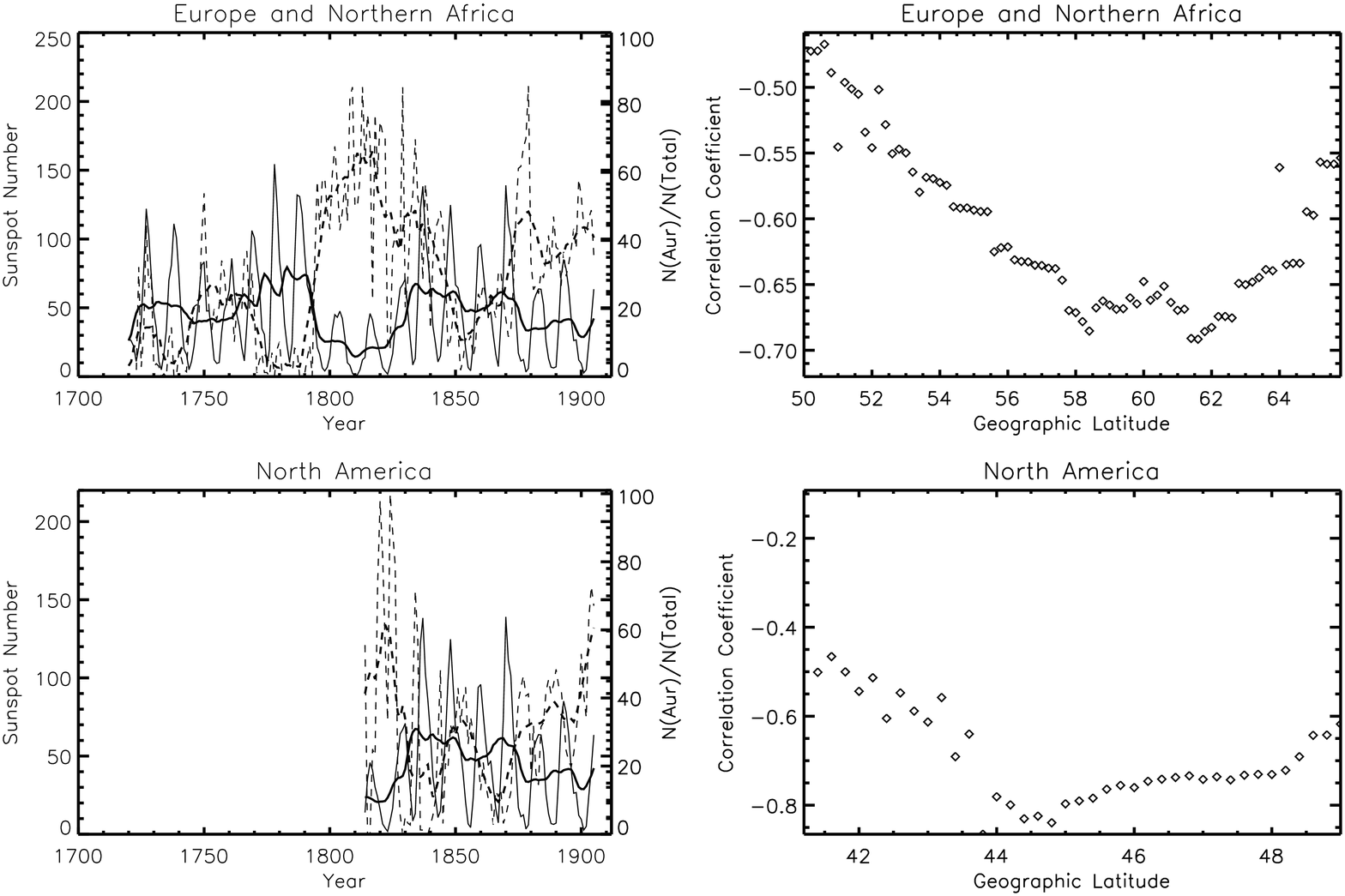}}
\caption{(Left-hand side): Relationship between the yearly sunspot number (solid line) and the 
fraction of aurorae observed above a critical latitude at which  the highest correlation is found (dashed line). 
The thicker line corresponds to an 11-year smoothing. (Right-hand side): The variation of the correlation coefficient 
for different critical latitudes.}
\label{GloLat6y}
\end{figure}

The best correlation was obtained at 61.4 degrees 
geographic latitude for Europe ({\it r} = -0.70; {\it p} $<$ 0.005) and 44.8 degrees
for North America ({\it r} = -0.84; {\it p} $<$ 0.01), both in phase with the
sunspot number (Figure \ref {GloLat6y}).

\section {The Geomagnetic Latitude}

Since the frequency of the aurorae is related to distance
from the magnetic pole, it is more appropriate to plot the
observations {\it versus} magnetic latitude.

We computed the temporal evolution of the geomagnetic
latitude for every observing location during the entire
1700\,--\,1905 period. It is common to define within the main geomagnetic field the geomagnetic latitude $\phi$ according 
to the expression  $\tan \phi=(\tan I)/2$ (see Stacey, 1992; Buforn, Pro, and Ud\'\i as, 2012). 
We obtained the magnetic inclination (I) from the global geomagnetic model {\it gufm1} 
(Jackson, Jonkers, and Walker, 2000), which can be applied in the period covered by this study. 
Therefore, the geomagnetic latitude has been calculated 
for each auroral record from the value of the magnetic inclination.

\begin{figure}[ht]
 \centerline{\includegraphics[width=0.94\textwidth,clip=]{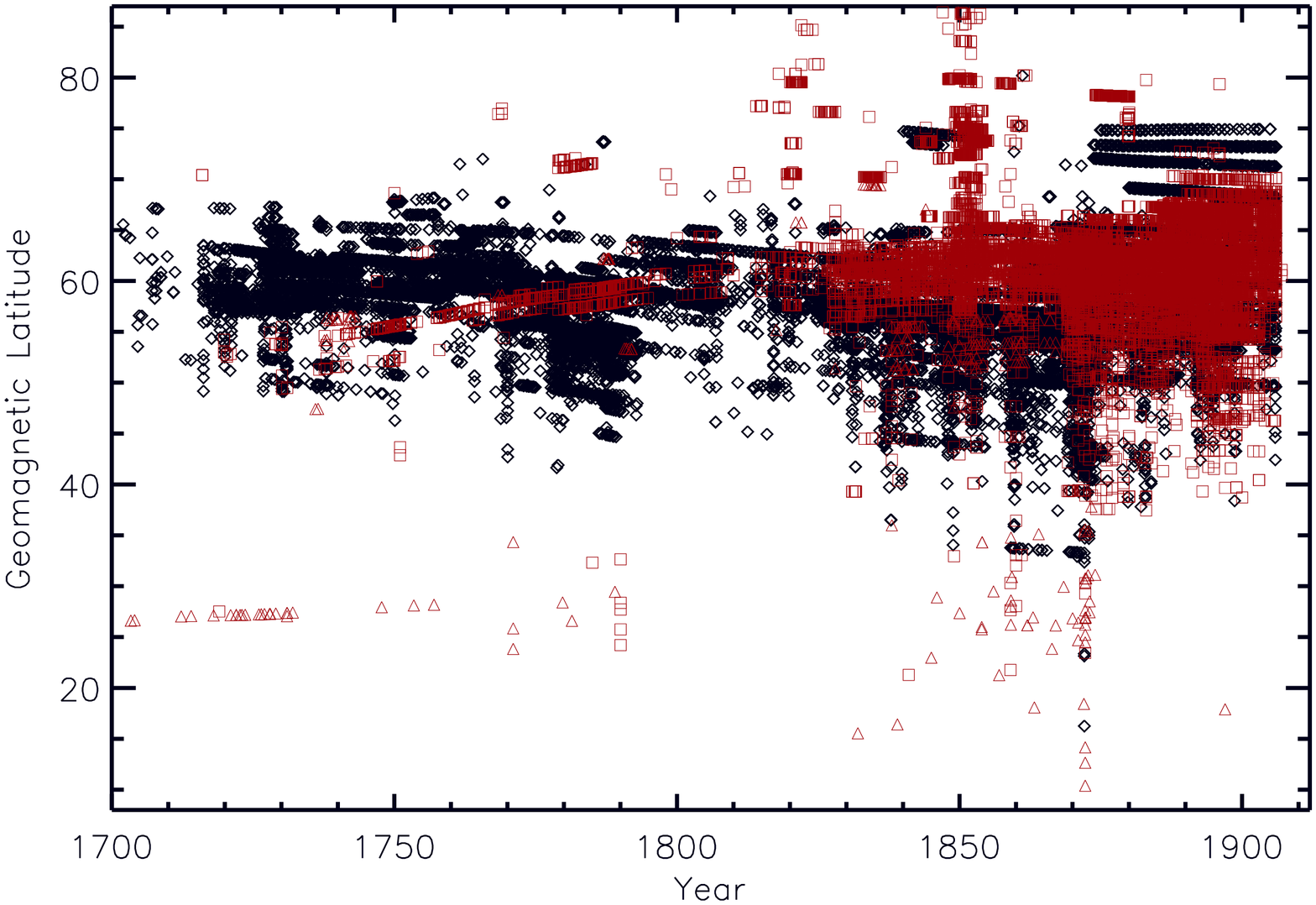}}
\caption{Latitude and time distribution of auroral events according to the magnetic latitude of the observing site. 
(Black diamonds) Europe and North Africa; (red squares) North America;  (red triangles) Asia.}
\label{MagLatGlo}
\end{figure}

We applied this procedure to all of the observing sites. Figure
\ref {MagLatGlo} shows the results for Europe, North
America, and Asia. One observes the Dalton Minimum between
the two high-activity episodes. However, the few more
southern magnetic latitudes during the 18$^{th}$ century are
mainly covered by Asian data. Two episodes with
high-latitude aurorae occur in 1820 and 1840 close to the
minimum phase of the corresponding solar cycles.  An unusual
aurora was visible close to the magnetic North Pole (Fort
Conger) on 17 November 1882. These observations was made by the expedition
commanded by A.W. Greeley from July 1882 to August 1883, in
the framework of the activities of the First Polar Year
(Taylor, 1981). Recently, Singh {\it
et al.}  (2012) have studied the characteristics of
high-latitude storms above the classical auroral oval.

\begin{figure}[ht]
 \centerline{\includegraphics[width=0.92\textwidth,clip=]{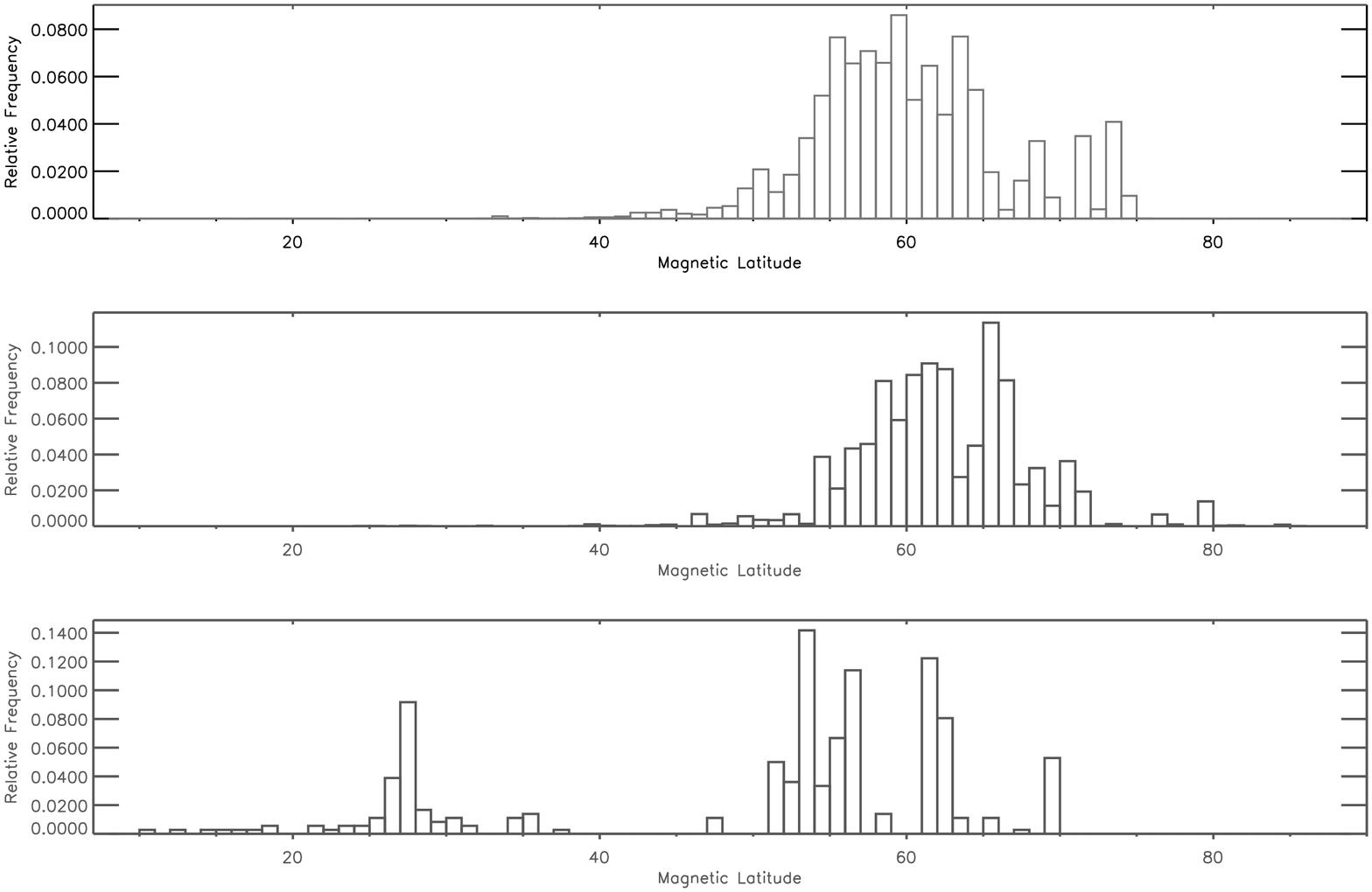}}
\caption{Histogram of the geomagnetic latitudes. (Top) Europe and North Africa; (middle) North America; (bottom) Asia.}
\label{GloMagLat}
\end{figure}

Figure \ref {GloMagLat} shows the histograms of the
geomagnetic latitudes for the three continental
masses. The major contribution of the
otherwise sparse Asian data to the low geomagnetic
latitudes stands out. As with the geographic latitudes, the geomagnetic
latitudes also have their absolute minima for the auroral
event of 1872.

\begin{figure}[ht]
 \centerline{\includegraphics[width=0.92\textwidth,clip=]{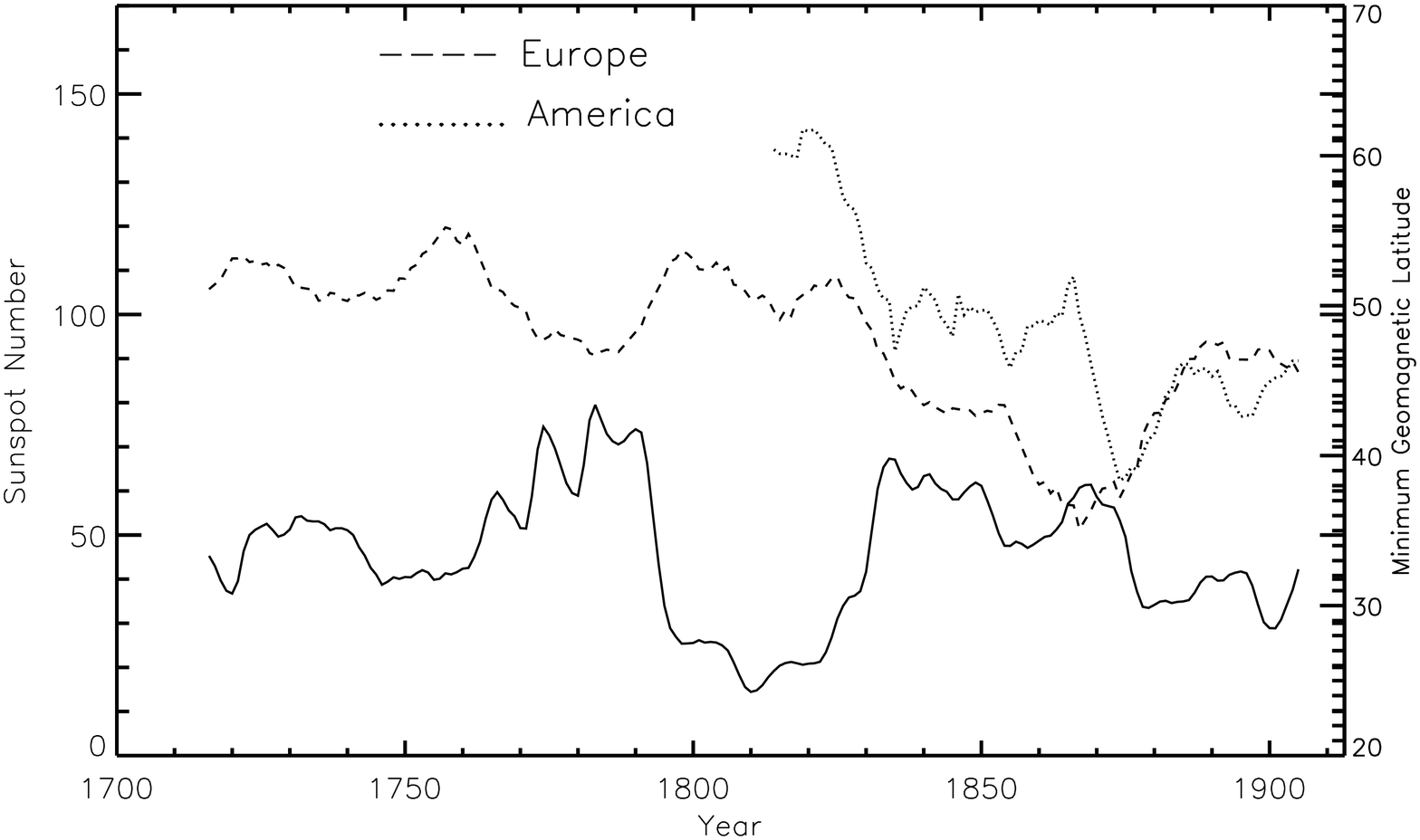}}
\caption{Variation of the annual minimum magnetic latitude -- dashed line: Europe (values after 1716); dotted line: 
America (Values after 1814) -- with the annual sunspot number (solid line). The data correspond to an 11-year smoothing.}
\label{GlMagLatMin}
\end{figure}

Figure \ref {GlMagLatMin} shows the temporal variation of the
yearly minimum geomagnetic latitude with the annual averaged
sunspot number. The anti-correlation is similar for Europe
and North America (-0.36 {\it vs} --0.32), but the level of significance is 
very low, especially for the American data. We must notice that the minimum
geomagnetic latitude shows a decrease during the first
half of the 19${th}$ century, most clearly visible for the American
data.

\begin{figure}[ht]
\centerline{\includegraphics[width=0.92\textwidth,clip=]{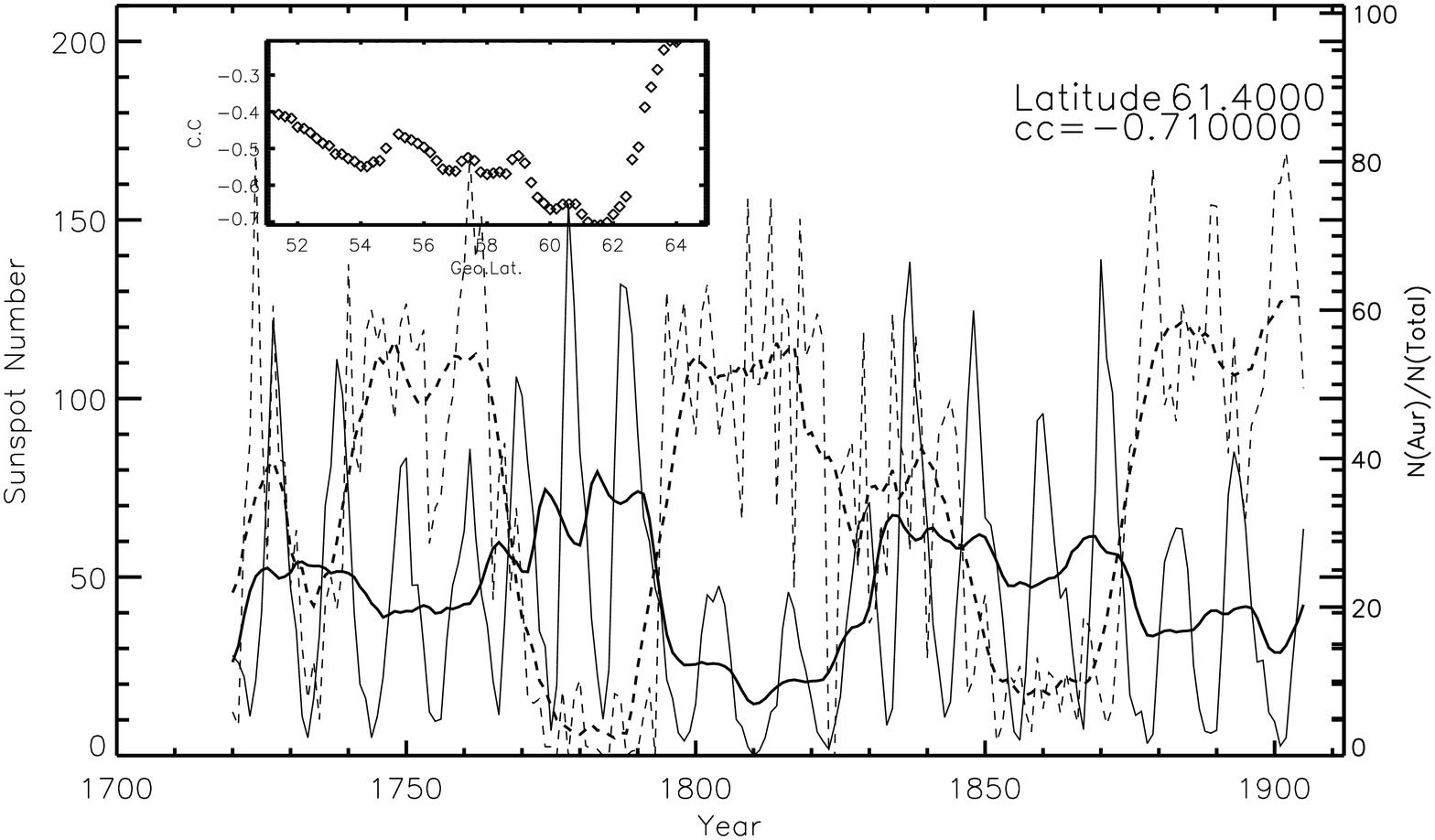}}
\caption{European and North African data: Correlation of the annual sunspot number with the fraction of 
auroral events above a critical geomagnetic latitude. The inset shows the variation of the correlation 
coefficient for different geomagnetic latitudes.}
\label{EurMagLatcrit}
\end{figure}

\begin{figure}[ht]
 \centerline{\includegraphics[width=0.92\textwidth,clip=]{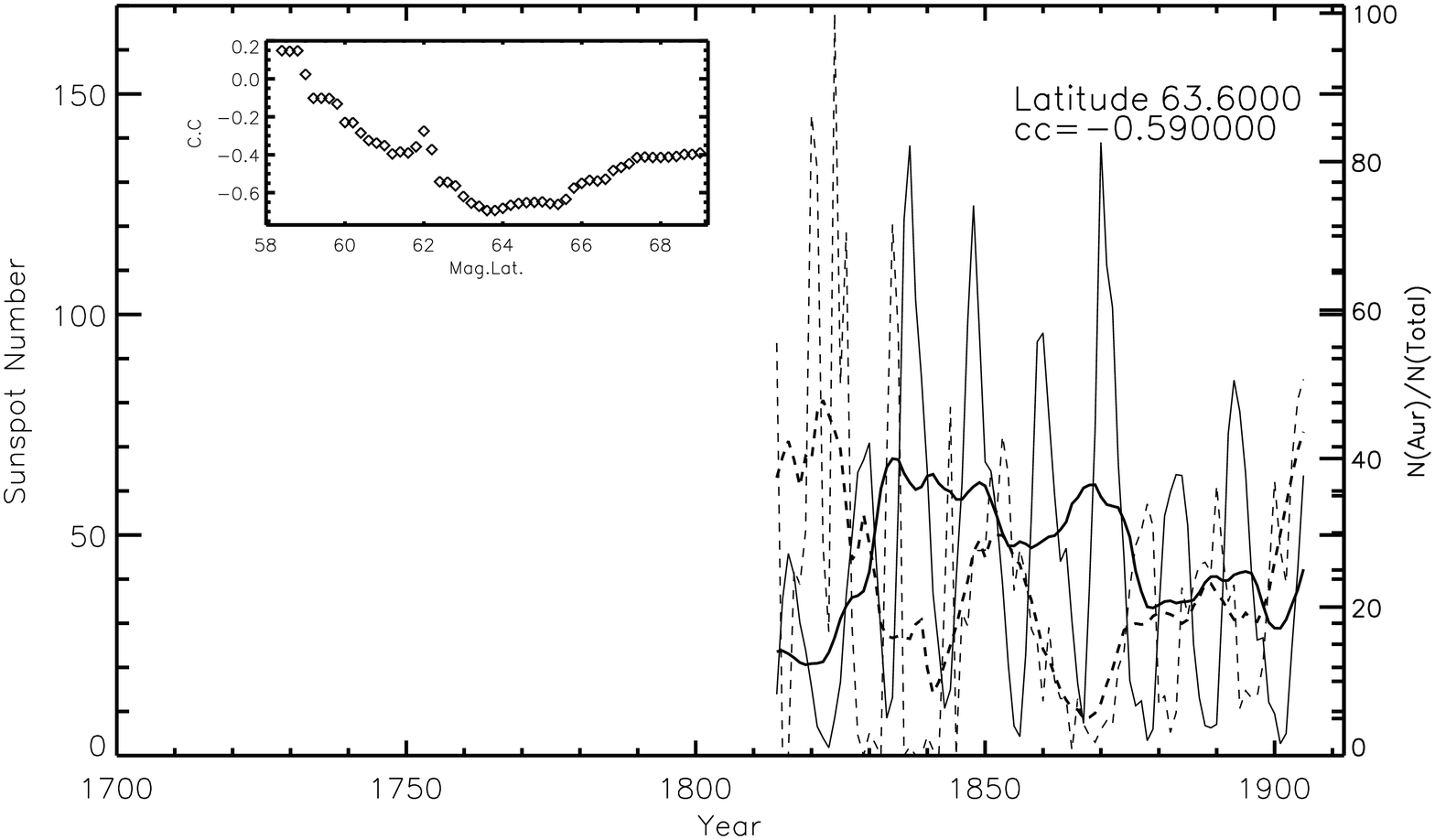}}
\caption{North America data: Correlation of the annual sunspot number with the fraction of auroral events 
above a critical geomagnetic latitude. The inset shows the variation of the correlation coefficient for 
different geomagnetic latitudes.}
\label{AmMagLatcrit}
\end{figure}

For the yearly number of aurorae above a critical magnetic
latitude (Figures \ref {EurMagLatcrit} and \ref
{AmMagLatcrit}), the best anti-correlation is found for a
magnetic latitude of around 61.4 and 63.4 for the European
and American samples, respectively (see the insets in the
figures for the variation of the correlation coefficients
with the geomagnetic latitude). As expected from previous results, the 
level of significance for this correlation was better for the Europe ({\it p} $<$ 0.005) than 
for the North America sample ({\it p} $<$ 0.05).

\begin{figure}[ht]
 \centerline{\includegraphics[width=0.92\textwidth,clip=]{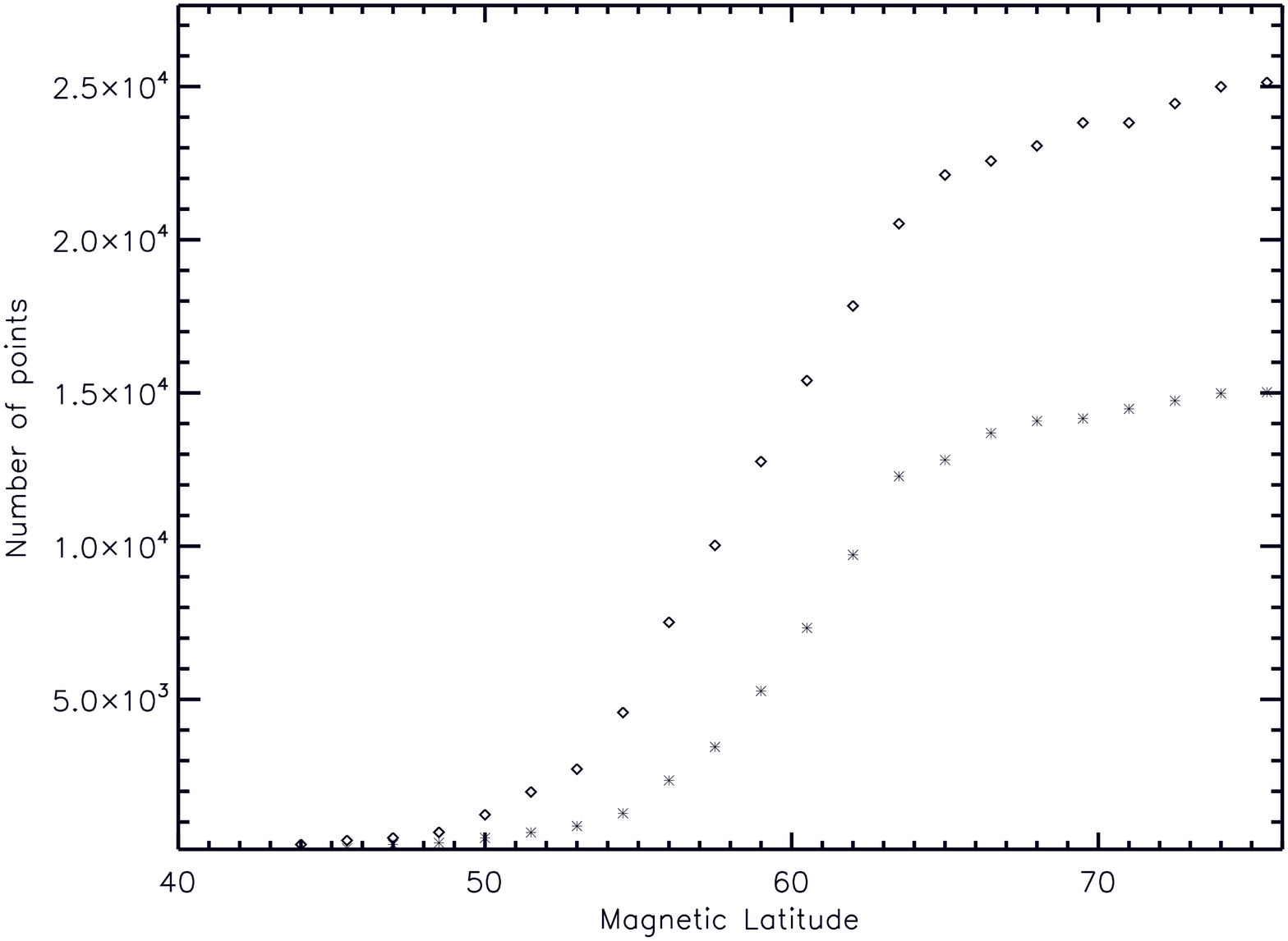}}
\caption{Cumulative values of the number of aurorae below a certain minimum geomagnetic latitude. ($\diamond$) 
European and N. African sample; ($\ast$) North American data.}
\label{GloMagLatCum}
\end{figure}

We repeated the calculation of the cumulative number of
auroral events, but now for the geomagnetic latitudes. The
results (Figure \ref {GloMagLatCum}) show the behaviour to
be similar for the two continents, although there is a lack
of high geomagnetic latitudes for the North American sample
(Canada and Alaska). Liritzis and Petropoulos (1987) already
noted a marked change at a geomagnetic latitude of 57\,--\,58
degrees.

 \section {Conclusions} 

We have studied the broad statistical properties (spatial
and temporal variation) of several catalogues and reports of
auroral observations during the 18$^{th}$ and 19$^{th}$ centuries. To
this end, we assigned geographic and geomagnetic
latitudes to more than 80\thinspace000 auroral observations.  For the
analysis, we divided the Northern Hemisphere into three
large continental masses corresponding to Europe, North
America, and Asia.

The sample presented numerous spatial and temporal gaps,
mainly evident in the North American data due to the
progressive European occupation of large areas outside the original
New England states. It was clear that further investigation
into low-latitude auroral events is important.  The
historical auroral records investigated can serve as a
proxy for reconstructing the long-term variability of the
heliosphere.

To gain insight into the broad pattern of the spatial and
temporal variations of auroral observations in the Northern
Hemisphere,  an 11-year smoothing procedure was applied to the
data to enhance this long-term component.

We suggested and applied a method to separate the
contribution of the different solar sources of auroral
events, which was based on the different slopes of the
cumulative numbers of auroral events per latitude bin.  The
low-latitude segment mainly corresponded to strong CMEs
taking place around the solar cycle maxima.  The middle
segment corresponded to standard CMEs, and the high-latitude
segment to high-speed streams of solar wind originating from
coronal holes. The limit between the last two sources is in
a narrow belt between 61 and 64 degrees of geomagnetic
latitude. One must bear in mind that the
geoeffectiveness of such transitory solar events depends
critically on the level of the IMF and its orientation with
respect to the Earth's magnetosphere.

The spatial distribution of the available data is clearly
determined by the populations living at high-latitude sites
which were clearly greater in Europe than in North America.
The same may apply to the low-latitude sites in the two
continents.  This masks the expected better correlation of
the auroral parameters with the geomagnetic than
with the geographic latitude.

The existence of several minima was clearly seen. In addition to
the Silverman and Dalton episodes, the discontinuity at the
beginning of the 20$^{th}$ century merits special attention.
This topic will be investigated in detail in a forthcoming
article.

\begin{acks}

The authors wish to express their gratitude to Sam Silverman
who has undertaken the immense task of collecting thousands
of auroral reports around the world and making them
available to the scientific community. Some well-known
auroral catalogues (Fritz, Angot) used for this study are
from Jack Eddy's Compilation of Auroral Catalogues and were
obtained from the Research Data Archive (RDA), which is
maintained by the Computational and Information Systems
Laboratory (CISL) at the National Center for Atmospheric
Research (NCAR). NCAR is sponsored by the National Science
Foundation (NSF).  The original data are available from the
RDA (dss.ucar.edu) with dataset number ds836.0.
Support from the Junta de Extremadura (Research Group Grant
No. GR10131) and the Ministerio de Econom\'ia y
Competitividad of the Spanish Government (AYA2011-25945) is
also gratefully acknowledged. Finally, we thank an anonymous
referee for useful suggestions.

\end{acks}

%%% %%%%%%%%%%%%%%%%%%%%%%%%%%%%%%%%%%%%%%%%%%%%%%%%%%%%%%%%%%%
%% Bibliography
%
% Using BibTeX
%
% \bibliographystyle{spr-mp-sola}
% %\bibliographystyle{spr-mp-sola-cnd} %% Alternative style: no title, no concluding page
% \bibliography{<bib file>}  

\begin{thebibliography}{}
% \bibitem[\protect\citeauthoryear{Author}{Year}]{key}
%   <bibliographical entry>
%
% \bibitem[\protect\citeauthoryear{}{}]{}



\bibitem[]{}
Angot, A.: 1897, 
\newblock {\em The Aurora Borealis}.
\newblock D. Appleton \& Co., New York.

Anonymous: 1883, 
\newblock {\em The periodicity of aurorae}.
\newblock {The Observatory} {\bf 6}, 116.

\bibitem[]{}
Aragon\`es Valls, E., Orgaz Gargallo, J.: 2010,
\newblock {\em Treb. Mus. Geol. Barcelona} {\bf 17}, 45.

\bibitem[]{}
Basurah, H.M.: 2004,
\newblock {\em Solar Phys.} {\bf 225}, 209.

\bibitem[]{}
Bernard, F.: 1837,
\newblock {\em Am. J. Sci. Arts} {\bf 34}, 267.

\bibitem[]{}
Bravo, S., Otaola, J.A.: 1990,
\newblock {\em Annales Geophysicae} {\bf 8}, 315.

\bibitem[]{}
Broughton, P.: 2002,
\newblock {\em J. Geophys. Res.} {\bf 107}, 1152. 

\bibitem[]{}
Buforn, E., Pro, C., Ud\'\i as, A.: 2012, 
\newblock {\em Solved Problems in Geophysics }, Cambridge University Press.

\bibitem[]{}
Chapman, S.: 1957,
\newblock {\it Ann. IGY} {\bf 4} 386.

\bibitem[]{}
Chernosky, E.J.: 1966,
\newblock {\em J. Geophys. Res.} {\bf 71}, 965.
 
\bibitem[]{}
Dalton, J.: 1873, 
\newblock {\em Meteorological Observations and Essays}.

\bibitem[]{}
Du, Z.: 2012,
\newblock {\em Solar Phys.} {\bf 278}, 203.

\bibitem[]{}
Eather, R.: 1980, 
\newblock {\em Majestic Lights: The Aurora in Science, History and the Arts}.
\newblock American Geophysical Union.

\bibitem[]{}
Feldstein, Y.I.: 1986,
\newblock {\em EOS Trans. AGU} {\bf 67}, 761.

\bibitem[]{}
Feyman, J., Silverman, S.: 1980,
\newblock {\em J. Geophys. Res.} {\bf 85, A6}, 2991.

\bibitem[]{}
Feyman, J., Gabriel, S.B.: 1990,
\newblock {\em Solar Phys.} {\bf 127}, 393.

\bibitem[]{}
Fritz, H.: 1873, 
\newblock {\em Verzeichniss Beobachteter Polarlichter}, C. Gerold's Sohn.

\bibitem[]{}
Georgieva, K., Kirov, B., Gavruseva, E.: 1967, 
\newblock {\em Phys. Chem. Earth} {\bf 31}, 81. 

\bibitem[]{}
Gnevyshev, M.N.: 1967, 
\newblock {\em Solar Physics},{\bf 1}, 107. 

\bibitem[]{}
Gonzalez, W.D., Tsuratani, B.T., Lepping, R.P., Schwenn, R.: 2002,
\newblock {\em J. Atmos. Solar-Terr. Phys.} {\bf 64}, 173.

\bibitem[]{}
Greeley: 1881, 
\newblock {\em Chronological List of Auroras observed from 1870 to 1879}, Prof. Pap. of the Signal Service, US Government Printing Office.

\bibitem[]{}
Groub\'e, W., Brigonnet, R., Geneslay, E., Moreau, J., Rebuffat, L., Saug\`ere, J.: 2006,
\newblock {\em L'Astronomie} {\bf 71}, 103.

\bibitem[]{}
Green, J.L., Boardsen, S.: 2006,
\newblock {\em Adv. Space Res.} {\bf 38}, 130.

\bibitem[]{}
Greenwood, G.: 1872,
\newblock {\em Nature} {\bf 5}, 400.

\bibitem[]{}
Halley, E.: 1716, 
\newblock {\it Phil. Trans. Roy. Soc. London} {\bf 24}, 406.

\bibitem[]{}
Harrison, G.: 2005, 
\newblock {\it Astron. Geophys.} {\bf 46}, 4.31.

\bibitem[]{}
Hecht, J.~H., Mulligan, T., Strickland, D.~J., Kochenash, A.~J.,
Murayama, Y., Tanaka, Y.~M., {\it et al.}: 2008,
\newblock {\it J. Geophys. Res.} {\bf 113}, A013010. 

\bibitem[]{}
Jackson, A., Jonkers, A.R.T., Walker, M.R.: 2000, 
\newblock {\em Phil. Trans. Roy. Soc. London A} {\bf 358}, 957.

\bibitem[]{}
Jones, H.S.: 1955, 
\newblock {\em Greenwich Observatory Sunspot and Geomagnetic Storm Data},
\newblock Her Majesty's Stationery Office.

\bibitem[]{}
Kimmball, D.S.: 1960,
\newblock {\it Scientific Report No. 6}, Geophysical Institute, University of Alaska.

\bibitem[]{}
Krivsk\'y, L.: 1996,
\newblock {\it Publ. Astron. Inst. Acad. Sci. Czech Republic} {\bf 75}.

\bibitem[]{}
Krivsk\'y, L., Pemjl, K.: 1988,
\newblock {\it Publ. Astron. Inst. Czechosl.
Acad. Sci.} {\bf 84}.

\bibitem[]{}
Kumar, P., Manoharan, P.K., Uddin, W.: 2010, 
\newblock {\em Geomag. Aeron.} {\bf 49}, 961.

\bibitem[]{}
Lang, A.S.: 1849,
\newblock {\em Mon. Not. Roy. Astron. Soc.} {\bf 9}, 148.

\bibitem[]{}
Lee, E.H., Ahn, Y.S., Yang, H.J., Chen, K.Y.: 2004,
\newblock {\em Solar Phys.} {\bf 224}, 373.

\bibitem[]{}
Legrand, J.P., Simon, P.A.: 1981,
\newblock {\em Solar Phys.} {\bf 70}, 173.

\bibitem[]{}
Legrand, J.P., Simon, P.A.: 1987,
\newblock {\em Annales Geophysicae} {\bf 5}, 161.

\bibitem[]{}
Le Mou\"el J.L., Blanter, E., Shnirman, M., Courtillot, V.: 2012, 
\newblock {\em J. Geophys. Res.} {\bf 117}, A09103.

\bibitem[]{}
Liritzis, Y., Petropoulos,B.: 1987,
\newblock {\em Earth, Moon, Plan.} {\bf 39}, 75

\bibitem[]{}
Livesey, R.L.: 1991,
\newblock {\em J. British Astron. Association} {\bf 101}, 233.

\bibitem[]{}
Lockwood, M., Stamper, R., Wild, M.N.: 1999,
\newblock {\em Nature} {\bf 399}, 437.

\bibitem[]{}
Lockwood, M.: 2003,
\newblock {\em J. Geophys. Res.} {\bf 108}, 1128.

\bibitem[]{}
Loomis, E.: 1860,
\newblock {\em Am. J. Sci. 2nd Ser.} {\bf 30}, 79.

\bibitem[]{}
Loomis, E.: 1861,
\newblock {\em Am. J. Sci. 2nd Ser.} {\bf 32}, 71.

\bibitem[]{}
Love, J.J.: 2011,
\newblock {\em Ann. Geophysicae} {\bf 29}, 1365.

\bibitem[]{}
Lovering, J.: 1866,
\newblock {\em Mem. Am. Acad. Arts Science} {\bf 10}, 9.

\bibitem[]{}
Lueders, F.G.T: 1984,
\newblock {\em Publ. Washburn Obs.} {\bf 12}, 20.

\bibitem[]{}
Mairan, J.J.: 1733,
\newblock {\em Trait\'e physique et historique de l\'e\, Aurore Bor\'eale},
\newblock Impremerie Royale.

\bibitem[]{}
Mendillo, M., Keady, J.: 1976,
\newblock {\em EOS Trans. AGU} {\bf 57}, 485.

\bibitem[]{}
Milan, S.E., Hutchinson, J., Boakes, P.D., Hubert, B.: 2009,
\newblock {\em Ann. Geophys.} {\bf 27}, 2913.

\bibitem[]{}
Moritz, H.: 1980,
\newblock {\em J. Geodesy} {\bf 74}, 128.

\bibitem[]{}
Moss, K., Stauning, P.: 2012,
\newblock {\em History of Geo- Spa. Sciences} {\bf 3}, 53.

\bibitem[]{}
Mursula, K., Martini, D., Karinen, A.: 2004,
\newblock {\em Solar Phys.} {\bf 224}, 85.

\bibitem[]{}
Nakazawa, Y., Okada, T., Shiokawa, K.: 2004,
\newblock {\em Earth Planets Space.} {\bf 56}, e41.

\bibitem[]{}
Nevanlinna, H., Kataja, E.: 1993,
\newblock {\em Geophys. Res. Lett.} {\bf 20}, 2703.

\bibitem[]{}
Ouattara, F., Amory-Mazaudier, C., Menvielle, M., Simon, P., Legrand, J.P.: 2009,
\newblock {\em Ann. Geophys.} {\bf 27}, 2045.

\bibitem[]{}
Rassoul, H. K., Rohrbaugh, R.P., Tinsley, B.A., Slater, D.W.: 1993,
\newblock {\em J. Geophys. Res.} {\bf 98, A5}, 7695. 

\bibitem[]{}
Reiff, P., Luhmann, J.G.: 1986,
\newblock In {\em Solar-wind magnetosphere coupling}, Y. Kamide (ed.), pp.453-476, D. Reidel.

\bibitem[]{}
R\'ethly, A., Berkes, Z: 1963,
\newblock In {\em Nordlicht-Beobachtungen in Ungarn}, Budapest, Ungarische Akademie der Wissenschaften.

\bibitem[]{}
Roach, F.E., Moore, J.G., Bruner, E.C., Cronin, H., Silverman, S.M.: 1960,
\newblock {\em J. Geophys. Res.} {\bf 65}, 3575.

\bibitem[]{}
Robbrecht, E., Berghmans, D., Van der Linden, R.A.M.: 2009
\newblock {\em Astrophys. J.}, {\bf 691}, 1222. 

\bibitem[]{}
Rouillard, A.P., Lockwood, M., Finch, I.: 2007,
\newblock {\em J. Geophys. Res.} {\bf 112}, A105103.

\bibitem[]{}
Rubenson, R.: 1882, 
\newblock {\em Catalogue des aurores bor\'eales observ\'ees en Su\`ede}, Kl. Sven. Vetenskapsakad. Handl. 18, 1.

\bibitem[]{}
Schr\"oder, W.: 1966,
\newblock {\em Gerl. Beitr. Geophys. } {\bf 75}, 436.

\bibitem[]{}
Schr\"oder, W.: 1980,
\newblock {\em Planet. and Space Science} {\bf 46}, 461.

\bibitem[]{}
Schr\"oder, W., Shefov, N.N., Treder, H.J.: 2004,
\newblock {\em Ann. Geophysicae} {\bf 22}, 2273.

\bibitem[]{}
Sheeley, N.R., Howard, R.A.: 1980,
\newblock {\em Solar Phys.} {\bf 67}, 189.

\bibitem[]{}
Silverman, S.: 1992,
\newblock {\em Rev. Geophys.} {\bf 30}, 333.

\bibitem[]{}
Silverman, S.: 1995,
\newblock {\em J. Atmos. Solar-Terr. Phys.} {\bf 57}, 673.

\bibitem[]{}
Silverman, S.: 2003,
\newblock {\em J. Geophys. Res.} {\bf 108}, 8011.

\bibitem[]{}
Silverman, S.: 2008,
\newblock {\em J. Atmos. Solar-Terr. Phys.} {\bf 70}, 1301.

\bibitem[]{}
Silverman, S., Cliver, E.W.: 2001,
\newblock {\em J. Atmos. Solar-Terr. Phys.} {\bf 63}, 523.

\bibitem[]{}
Singh, A.K., Sinha, A.K., Rawat, R., Jayashree, B., Pathan, B.M., Dhar, A.: 2012,
\newblock {\em Adv. Space Res.} {\bf 50}, 1512.

\bibitem[]{}
Siscoe, G.L.: 1980,
\newblock {\em Rev. Geophys.} {\bf 18}, 647.

\bibitem[]{}
Solanki, S., Sch\"ussler, M., Fligge, M.: 2002,
\newblock {\em Astron. Astrophys.} {\bf 383}, 706.

\bibitem[]{}
Solanki, S., Usoskin, I.G., Kromer, B., Sch\"ussler, M., Beer, J.: 2004,
\newblock {\em Nature} {\bf 431}, 1084.

\bibitem[]{}
Srivistava, N.J.: 2005,
\newblock {\em Ann. Geophys.} {\bf 23}, 2969.

\bibitem[]{}
Stauning, P.: 2011,
\newblock {\em Hist. Geo. Space Sciences} {\bf 2}, 1.

\bibitem[]{}
Stacey, F.D.: 1992, 
\newblock {\em Physics of the Earth, third edition},Brookfield Press.

\bibitem[]{}
Stone, E.J.: 1872,
\newblock {\em Nature} {\bf 5}, 443.

\bibitem[]{}
Svalgaard, L., Cliver, E.W., Le Sager, P.: 2004,
\newblock {\em Adv. Space Res.}, {\bf 34}, 236.

\bibitem[]{}
Taylor, C.J.: 1981,
\newblock {\em Artic} {\bf 34}, 376.

\bibitem[]{}
Tinsley, B.A., Rohrbaugh, R., Rassoul, H., Sahai, Y., Teixeira, N.R.: 1986,
\newblock {\em J. Geophys. Res.} {\bf 91}, 11257.

\bibitem[]{}
Traversi, R., Usoskin, I.G., Solanki, S.K., Becagli, S., Frezzetti, M., Severi, M., Stenni, B., Udisti, R.: 2012,
\newblock {\em Solar Phys.} {\bf 280}, 237.

\bibitem[]{}
Tromholt, S.: 1898, 
\newblock {\em Catalogue der in Norwegen bis Juni 1872 beobachteter Nordlichter}, Jacob Dybwad Forlag.

\bibitem[]{}
Tsurutani, B.T., Gonzalez, W.D., Gonzalez, A.L.C., Guarnieri, F.L., Gopalswamy, N., Grande, M.,
Kamide, Y., Kasahara, Y., Lu, G., Mann, I., McPherron, R., Soraas, F. and Vasyliunas, V.: 2006
\newblock {\em J. Geophys. Res.}, {\bf 111}, A10, 10.1029/2005JA011273 

\bibitem[]{}
Usoskin, I.G., Mursula, K., Solanki, S.K., Sch\"ussler, M., Kovaltsov, G.A.: 2002,
\newblock {\em J. Geophys. Res.} {\bf 107}, 1374.

\bibitem[]{}
Usoskin, I.G., Kovaltsov, G.A.: 2012,
\newblock {\em Astrophys. J.} {\bf 757}, 92.

\bibitem[]{}
Vallance Jones, A.: 1992,
\newblock {\em Can. J. Phys.} {\bf 70}, 479.

\bibitem[]{}
Vaquero, J. M., Gallego, M.C., Garcia, J.A.: 2003,
\newblock {\em J. Atmos. Solar-Terr. Phys.} {\bf 65}, 677.

\bibitem[]{}
Vaquero, J. M., Trigo, R.M.: 2005,
\newblock {\em Solar Phys.} {\bf 231}, 157.

\bibitem[]{}
Vaquero, J. M., Trigo, R., Gallego, M.C.: 2007,
\newblock {\em Earth, Planets, Space} {\bf 59}, e49.

\bibitem[]{}
Vaquero, J. M., Valente, M.A., Trigo, R.M., Ribeiro, P., Gallego, M.C.: 2008,
\newblock {\em J. Geophys. Res.} {\bf 113}, A08230.

\bibitem[]{}
Vaquero, J. M., Gallego, M.C., Barriendos, M., Rama, E., S\'anchez Lorenzo, A.: 2010,
\newblock {\em Adv. Space Res.} {\bf 45}, 1388.

\bibitem[]{}
Vaquero, J.M., V\'azquez, M.:2009, 
\newblock {\em The Sun recorded through History}, Springer.

\bibitem[]{}
V\'azquez, M., Vaquero, J.M., Curto, J.J.: 2006,
\newblock {\em Solar Phys.} {\bf 238}, 405. 10.1007/s11207-006-0194-2

\bibitem[]{}
V\'azquez, M., Vaquero, J.M.: 2010,
\newblock {\em Solar Phys.} {\bf 267}, 431. 10.1007/s11207-010-9650-0

\bibitem[]{}
Vennerstr$\o$n, S., Friis-Christensen, E.: 1996,
\newblock {\em J. Geophys. Res.} {\bf 101}, 24727.

\bibitem[]{}
Verbanac, S., Vrsnak,B., Veronig, A., Temmer,M.: 2011,
\newblock {\em Astron. Astrophys.} {\bf 526}, A20.

\bibitem[]{}
Visser, S.W.: 1942,
\newblock {\em Catalogue of Dutch Aurorae 1728 - 1940} (rda.ucar.edu/datasets/ds836.0/)

\bibitem[]{}
Webb, D.F., Howard, R.A.: 1994,
\newblock {\em J. Geophys. Res.} {\bf 99}, 4201.

\bibitem[]{}
Willis, M., Stephenson, F.R., Fang, H.: 2007,
\newblock {\em Ann. Geophysicae} {\bf 25}, 417.

\bibitem[]{}
Yau, K.K.C., Stephenson, F.R., Willis, D.M.: 1995, 
\newblock {\em A catalogue of auroral observations from China, Korea and Japan (193 B.C. - A.D. 1770)}, 
Rutherford Appleton Lab.

\end{thebibliography}
%
% Without BibTeX 

\end{article}
\end{document}